\documentclass[showpacs,superscriptaddress,preprint,prl]{revtex4}
\usepackage{amsmath,graphicx}

\begin{document}            

\title{  Signature of  gate-tunable  magnetism in graphene grafted with  Pt-porphyrins}
\author{Chuan Li $^{1}$, Katsuyoshi Komatsu $^2$, S. Bertrand $^{1}$, G. Clav\'e$^3$, S. Campidelli$^3$, A. Filoramo$^3$, S. Gu\'eron$^{1}$ and H. Bouchiat$^{1}$ }
\affiliation{$^1$ LPS, Univ. Paris-Sud, CNRS, UMR 8502, F-91405 Orsay Cedex, France,$^2$ WPI Center for Materials Nanoarchitechtonics (WPI-MANA), National Institute for Materials Science (NIMS), Tsukuba, Ibaraki 305-0044, Japan,$^3$ CEA Saclay, IRAMIS, NIMBE (UMR 3685), Laboratoire d\rq{}Innovation en Chimie des Surfaces et Nanosciences (LICSEN), F-91191 Gif sur Yvette, France }

\begin{abstract}
Inducing magnetism in graphene holds great promises, such as controlling the exchange interaction with a gate electrode, and generating exotic magnetic phases.  Coating graphene with magnetic molecules or atoms has so far mostly  lead to decreased graphene mobility. In the present work, we show that Pt-porphyrin molecules adsorbed on graphene lead both to an enhanced mobility, and to gate-dependent magnetism.  We report that porphyrins can act both as donor or acceptor molecules, depending on the initial doping of the graphene sheet. The porphyrins transfer charge and  ionize around the charged impurities on graphene, and,  consequently,  the graphene doping is decreased and its mobility is enhanced.

In addition,  ionized porphyrin molecules carry a magnetic moment. Using the sensitivity of mesoscopic transport to magnetism, in particular the superconducting proximity effect and conductance fluctuations, we explore the magnetic order induced in graphene by the interacting magnetic moments of the ionized porphyrin molecules.

Among the signatures of magnetism, we find two-terminal-magnetoresistance fluctuations with an odd component, a tell-tale sign of time reversal symmetry breaking at zero field, that does not exist in uncoated graphene samples. When graphene is connected to superconducting electrodes, the induced magnetism leads to a gate-voltage-dependent suppression of the supercurrent, modified magnetic interference patterns, and gate-voltage-dependent magnetic hysteresis. The magnetic signatures are greatest for long superconductor/graphene/superconductor junctions, and for samples with the highest initial doping, compatible with a greater number of ionized, and thus magnetic porphyrin molecules.  Our findings suggest that long-range (of the order of the coherence length, or micrometers) magnetism is  induced through graphene by the ionized porphyrins' magnetic moment. This magnetic interaction is controled by the density of carriers in graphene, a tunability that could be exploited in spintronic applications.

\end{abstract}
 
\maketitle
\section{Introduction} 
Because of its conical band structure 
and the possibility to  continuously tune the Fermi level with a gate voltage,  graphene has opened a  broad new field of investigation of 2D electronic transport.  At low temperature  the phase coherence length, of the order of one micron, offers the possibility to  explore mesoscopic aspects of  transport in graphene such as conductance fluctuations and proximity-induced superconductivity.  More than classical transport, quantum transport is especially sensitive to the nature of scattering impurities on graphene,  in particular to their charged, neutral or polar character \cite{imp}.  The spatial  extent of the scattering potential  also plays an important role: short-range  neutral scatterers cause intervalley scattering,  in contrast to long range charged impurities. Beyond the characterization of intrinsic impurities of graphene, it is tempting to add specific impurities to induce new  functionalities and tune them in a controlled way.   Adsorbates on graphene transistors have for instance been shown to affect the transport characteristics of graphene-based sensors\cite{sensors}. On a more fundamental level, a longstanding goal is to
induce  and control physical properties involving spin degrees of freedom such as spin-orbit coupling and magnetism. Specific signatures of  spin-orbit or magnetic scatterers are expected, whose effect should be tunable with the gate voltage \cite{castro08,Guinea12,Alicea,Asmar}.    
To reach this goal, graphene has been coated with magnetic  (transition metal or rare earth)  atoms \cite{LauKawakami,Girit} as well as molecular magnets, with mixed results. Adsorbed magnetic atoms reduce graphene\rq{}s mobility, with no  clear concurrent magnetic signature. In contrast, the magnetisation reversal of molecular magnets has been detected in a graphene nanoconstriction \cite{Candini}.  Possible signatures of enhanced spin-orbit interactions have also been observed in graphene  coated with small non magnetic metallic clusters \cite{Ozyilmaz}. Magnetism could  also be induced by creating vacancies in graphene  \cite{Nair,vacancies}  or adsorbing atomic hydrogen \cite{McCreary}, with a doping-dependent magnetic signature.   However  in both cases this magnetism is induced at the cost of an unavoidable decrease of sample mobility.

In the present work, we functionalize graphene with  a thin film of Pt-porphyrins.  The  first layer directly  in contact with graphene can interact with graphene\rq{}s delocalised $sp^2$ orbitals and is expected to form an ordered array \cite{mayne}. Neutral Pt-porphyrins are non magnetic, but the ionized form carries a magnetic moment of one Bohr magneton \cite{ESR}. At room temperature we find that the molecules dope the graphene, demonstrating that charge transfer occurs.    Either electrons or holes can be transferred to  the graphene layer, depending on its initial doping. More surprisingly, the graphene\rq{}s mobility increases upon molecules deposition. Using  the high sensitivity of mesoscopic transport,  in particular the superconducting proximity effect and  conductance fluctuations \cite{lundberg}, we show evidence of  long range magnetism induced in several  porphyrin-coated  samples. This magnetism leads to field asymmetry of conductance fluctuations in samples with normal electrodes,  and suppression of supercurrent or modified Fraunhoffer patterns in samples with superconducting electrodes. 
The  signatures are largest for samples with high initial doping, for which more porphyrins ionize and thus become magnetic.

  One implication of these findings is that it is possible to control the number of magnetic porphyrins by their degree of ionization. In all experiments  this magnetism is found to depend on gate voltage, a tunability that  could be exploited in spintronic devices.  
 These findings also constitute  evidence for  Fermi-level controlled exchange interaction between localized spins and graphene.


\section{Sample preparation}
The samples are prepared  by exfoliation of high quality graphite and deposition onto an oxidized doped silicon wafer (acting as a backgate). Monolayers are selected by optical microscopy. The metallic contacts are made by electron beam lithography followed by  deposition of different metallic bilayers, Ti/Au, Ti/Al or Pd/Nb. All samples were measured at room temperature before deposition of the molecules.  The Pt-porphyrin, see Fig.1, contains a Pt atom at the center of the characteristic cyclic  organic cage of four pyrrole subunits interconnected via methine bridges (=CH). The porphyrins were prepared as described in \cite{prepaporphyrin1}. The electronic structure of porphyrins, as determined by optical absorption measurements and Scanning Tunneling  Spectroscopy, is characterized by a HOMO (Highest Occupied Molecular Orbital) -LUMO (Lowest Unoccupied Molecular Orbital) gap of the order of 2 eV \cite{HomoLumo}. We  deposited Pt porphyrins    at room temperature according to the following  protocol. We first  checked that deposition of the sole solvent   tetrahydrofuran (THF) does not  modify the gate voltage dependence of the sample's conductance. We  then deposited a  10 $\mu l$  drop of a  1 mM solution of  Pt  porphyrins in THF. This corresponds to a few hundred layers of porphyrins covering the graphene  after evaporation of the THF solvant. We have changed this number of layers by a factor 10 and find that the low temperature results (that demonstrate the gate-dependent magnetism) do not depend on the layer thickness, consistent with a porphyrin/graphene interaction restricted to the first layer.
We have functionalised over twenty graphene samples. Among those samples, ten were investigated at low temperature  both before and after functionalisation, and their characteristics are detailed in table 1.

\section{ Charge transfer between graphene and  porphyrins: Neutralization of graphene by porphyrins}

At room temperature (RT), we  systematically measured the gate  voltage dependence of the resistance  before and after porphyrin deposition. A striking neutralization effect, shown on Fig.1, was observed  on all investigated samples  :
the Dirac point of all  samples  is shifted to nearly zero gate voltage after deposition. 
This means not only that charges are transferred between porphyrins and graphene, but also that porphyrins can be both electron donors or acceptors. Since the HOMO-LUMO gap of the porphyrin (2 eV)  is much greater than the typical Fermi energy differences between the graphene samples we have investigated (0.1 eV), this  implies a local pinning of the  HOMO (LUMO) level of the porphyrins to the hole (electron) doped Fermi level of graphene. Electron transfer has  already been  reported for  highly hole-doped porphyrin-grafted carbon nanotubes\cite{Prato,Hecht}, and, more recently, for highly hole-doped  Zn-porphyrins-grafted graphene\cite{apl}.  However, the possibility to also inject holes in graphene with porphyrins has,  to our knowledge, not  previously been demonstrated.   

At room temperature the gate dependence of the resistance is hysteretic, with a slow exponential relaxation of the sample's resistance in response to a fast gate voltage change. We relate the RT hysteresis and slow relaxation to hopping processes through neighboring molecules\cite{Checcolia,Sedghi}, leading to a slow (hundred second time scale) charge transfer across  porphyrin layers above the graphene. 
 Fig.2 illustrates how the Dirac point is shifted to $-V_0$, with a broadened $R(V_g)$ curve, after keeping the sample at a non-zero gate voltage $V_0$  at room temperature. This phenomenon can be seen as temporary, artificial doping of graphene by the transfer and storage of charges in the molecules. Similar effects have been observed with graphene samples functionalized with  insulating nanoparticles (iron and titanium oxide, CdSe) showing evidence of  charge neutralization of graphene,  mobility increase  and gate  voltage dependent charge transfer between graphene and nanoparticles \cite{nanopart}.

When cooling the sample at zero gate voltage, the Dirac point stays unchanged close to $V_g=0$. 
In contrast to the RT behavior, there is no gate-dependent charge transfer at low T (below 4.2 K). This is demonstrated by magnetoresistance measurements in the quantum Hall  regime, for which 
  the carrier density $n_c$, estimated with the same parameters as  for the bare graphene sample (same capacitance, etc.) yields the Quantum Hall plateaus  for the coated sample at exactly the same  filling factors as before, see Fig. 3. 
Another important feature is the increase in sharpness of  the $R(V_g)$ curve  after porphyrin deposition, with a higher resistance at the charge neutrality point (Dirac point) (see Fig.3a), implying that coating with Pt-porphyrins results in a higher sample  mobility (from $ \mu = 8000$ to $10000 ~ cm^2 /(V s) $ near the Dirac point at $T=$ 4.2 K). Hall plateaus are also better defined. 
These observations,  along with the  shift of the Dirac point to $V_g\simeq 0$,  prove that  porphyrins ionize and neutralize charged impurities on graphene or on the silicon substrate, and therefore   decrease the disorder scattering.


This neutralization of graphene by ionization of the porphyrins  is concurrent with the formation of singly occupied  impurity states close to the Dirac point (distributed between the initial and final Dirac points). Whether or not these impurity states generate  a magnetic moment, corresponding  to the spin of the unpaired electron delocalized over the molecule, will depend on the energy of the impurity state, as well as on graphene's  Fermi level, and thus gate voltage, as described by Uchoa et al.\cite{castro08}.
 In the following we present signatures of this magnetic moment and its gate-voltage dependence, detected via phase coherent transport measurements at low temperature.

\section{Signature of magnetism on samples with normal contacts}

We first discuss 3 graphene samples, on the same Si substrate, with normal (non superconducting) Ti(6nm)/Au(100nm) contacts . Two of them (Au1, Au2) were coated with porphyrins, another (Au3) not. We find (see Fig. 4) that at 1 K the two-terminal resistance in perpendicular field of the uncoated sample exhibits reproducible mesoscopic  magneto-resistance  fluctuations due to interference between all the coherent trajectories across the samples. The amplitude of the conductance fluctuations is of the order of the conductance quantum. As for non magnetic mesoscopic samples, including graphene \cite {folk, Ojeda2010}, the curves of the uncoated samples are even  functions of magnetic field, as expected for 2 probes measurements on non magnetic mesoscopic samples that obey time reversal symmetry  in zero field (see Fig.4). By contrast, the magnetoresistance of the coated samples Au1 and Au2    contains an odd component, whose amplitude is of the order of  1/3 of the even component. This indicates that time reversal symmetry is broken in those samples in zero magnetic field, a fact that we attribute to  macroscopic magnetic moments perpendicular to the sample plane, extending on spatial scales comparable to the phase coherence and thermal lengths. Similar effects have been observed in mesoscopic spin glasses \cite{spinglass}. This  behavior must stem from  frozen, long range magnetic  correlations within the ionized  porphyrins, whose physical origin we discuss further in the article. 
 However, as shown in Fig.5, this  asymmetry of the magnetoresistance in a perpendicular  field is undetectable at very low temperature (100mK). Instead,  we find  a strong   asymmetric and hysteretic  magnetoresistance in parallel field, with a large odd component whose amplitude and sign   depend on gate voltage (see supplementary materials Fig.S1).  This asymmetric magnetoresistance in parallel field is not detectable at 1K.  Thus, it appears that the porphyrins\rq{} magnetic moment rotates from  in-plane at 100 mK to nearly out-of-plane at 1K. Since the average distance between ionized porphyrins is a few nanometers, our findings  suggest  a  relatively long-ranged magnetic interaction mediated by graphene\rq{}s conduction electrons or holes.

\section{Samples with superconducting contacts}

We now turn to samples with superconducting electrodes (S/Graphene/S  junctions), to exploit  the  sensitivity of  the Josephson current to magnetism. Whereas  small effects are found in short junctions  of length  $L$ smaller than the superconducting coherence length $ \xi_S$, (see samples Al1, Al2 and Al3 in supplementary materials), the most spectacular signatures  of induced magnetism   occur in long junctions for which $L\gg \xi_S $ or  equivalently $ \Delta \gg E_{Th}$ (with $\Delta$  the superconducting gap, $E_{Th} =\hbar D/L^2$  the Thouless energy and D the diffusion coefficient). 
This long junction regime can be achieved either with large gap  electrodes such as Nb (sample Nb) or with very disordered  samples with small $D$  (Ald1 and Ald2).

We first present  a 1.2 $\mu m$ long graphene sample connected to   Pd/Nb  superconducting electrodes (8nm Pd/70nm Nb, sample Nb),  with $ \xi_S =$ 30 $nm$. 
Fig. 6 compares the gate  voltage dependence of the differential resistance, at 100 mK, before and after  porphyrins deposition.  The resistance of the  uncoated graphene junction is zero at low dc current,  in highly doped regions (away from the Dirac point), both for hole and electron doping (Fig. 6a) \cite{Heersche},\cite{Komatsu12}:  this is the signature of a Josephson effect, a supercurrent running through the graphene because of the superconducting contacts.  We previously reported that the amplitude of the maximal supercurrent, defined as the junction\rq{}s critical  current, is strongly depressed in the region of the Dirac point. We attribute that effect,  observed only in long junctions, to specular reflections of the Andreev pairs on very low carrier density regions (puddles) where the local  Fermi energy is smaller than the proximity induced superconducting gap (see \cite{Komatsu12} for a detailed discussion of these findings). 
It is clear that at high doping ($V_g>20~V$) the supercurrent exists for both electron and hole doping, as also shown in fig 6c. In stark contrast to this {\it bipolar} Josephson effect through pristine graphene (Fig. 6a and 6c), we find that after deposition of the porphyrins, the Josephson current is enhanced in the hole doped region but suppressed in the electron-doped region. Only a dip in the differential  resistance is visible at low current  and high positive gate voltage, but no strictly zero resistance  and no supercurrent(Fig. 6b). Fig. 6c, which displays the critical current amplitude as a function of gate voltage, shows  this strong asymmetry of the gate voltage dependence of the supercurrent after deposition of the porphyrins. The absence of supercurrent  in the electron doped region was checked for gate voltages up to 60V.  

 We attribute this extinction of the critical current, at high  positive gate voltage only, to  an inhomogeneous magnetic field on graphene  created by the  staggered magnetism of the porphyrins.  This destroys the  constructive interferences between the  Andreev pairs 
that carry  the  Josephson current. This sensitivity of the critical current to small magnetic perturbations is illustrated by  the field dependence of the supercurrent in this junction, and its narrow  \cite{Fraunhoffer} interference Fraunhoffer pattern (Fig.6d).  It is clear that a field as small as a fraction of a Gauss can suppress the induced supercurrent through graphene, because of destructive interference between Andreev pairs diffusing across the graphene \cite{Chiodi}. This explains how the porphyrin's magnetic spins, if they lead to correlated  magnetic regions at positive gate voltage, can create an inhomogeneous magnetic flux sufficient to destroy the proximity effect, thereby leading to the observed unipolar supercurrent.   

 Another indication of gate-voltage -dependent magnetism comes from the magnetoresistance in a perpendicular magnetic field,  (Fig. 6e),  which is a hysteretic function of field at $V_g > 0$ , whereas there is no hysteresis for $V_g <0$. This absence  of magnetic hysteresis and the large Josephson current in the hole doped region   consistently indicate either a quenched magnetic moment for the porphyrins or a reduced exchange interaction between porphyrin\rq{}s localised spins and the graphene holes. Moreover the magnetic hysteresis  in the electron  doped region ($V_g > 0$) points to the formation of a magnetic order of the molecular spins, with   partially oriented magnetic domains, generating a non uniform magnetic field that is revealed by the magnetoresistance of graphene, and explains the suppressed Josephson current for this doping.  No such effect is visible at high hole doping, which we attribute to the absence of magnetism in this gate voltage range. 
The asymmetry of induced magnetism with respect to the sign of gate voltage is striking. It was not  observed  in samples Au1 and Au2 discussed in  the previous section but exists in samples with  TiAl contacts discussed in the next section. This gate voltage \lq\lq{}polarity'' is probably strongly dependent on the initial doping and mobility of the graphene before deposition of porphyrins, (see discussion below).

 The signatures of magnetism are even more striking  in  data obtained on  two last samples, Ald1 and Ald2, with  Ti/Al contacts.  These samples  were initially highly  electron-doped (Dirac point  below  $-40~V$, see inset of Fig. 7), as the result of  oxygen plasma cleaning of the substrate  {\bf before} graphene deposition.  Coating  with porphyrins resulted in a shift of graphene\rq{}s Dirac point  to nearly zero gate voltage, despite the strong  initial doping. This spectacular observation attests the huge capacity of charge transfer and neutralization of porphyrins. In this case, the porphyrins must  act as acceptors of electrons  to neutralize these initially electron-doped graphene samples.  We expect a much higher concentration of ionized porphyrins in these samples compared to the ones discussed above, where the Dirac point shift did not exceed 15 V. The high  resistance at the Dirac point, (40 $k\Omega$), indicates a mobility at least 10 times smaller  than the samples discussed previously. As a result we estimate a Thouless energy of the order of $4~\mu eV$,  much smaller than the superconducting contacts gap of the order of $200  ~\mu eV$. The samples are thus also in the long junction limit. We do not observe a  supercurrent, at any gate voltage, but see up to a $75\%$ drop of differential resistance at zero bias at high electron or hole doping. The low field magnetoresistance of sample Ald1 for different gate voltages is shown in Fig.7 .  Whereas  at high doping the magnetoresistance dips at zero field and displays oscillations typical of a Fraunhofer interference  pattern, the magnetoresistance is clearly peaked at $B=0$ close to the Dirac point, and displays jumps at specific magnetic fields.   Field asymmetry and hysteresis are observed for all gate voltages, indicating that this magnetoresistance is due to  correlated magnetic regions in the graphene sample.   This magnetism  is however  asymmetric with respect to the Dirac point. It is maximum for negative gate voltage and decreases for positive gate voltage. This is illustrated in the inset of Fig.7 showing the   variance $\delta R_{AS}/R $ of the odd component of the magnetoresistance renormalized to the average total resistance as a function of the gate voltage. We attribute this asymmetry (opposite to Nb sample case) to the initial strong electron doping of the samples.
We have conducted similar investigations  in short junction samples  (see supplementary materials).  Whereas porphyrins coating does not suppress the critical current   at zero  field, we do find  distortions of the Fraunhofer patterns, as well as  important  asymmetries with respect to the electron/hole doping  in finite magnetic field (at the edges  of the Fraunhofer lobes).

\section{Discussion}

We attribute the doping-dependent asymmetries,  hysteresis in magnetic field, and doping-dependent suppression of Josephson current, to a gate-voltage dependent magnetism due to the ionized  porphyrin molecules. Moreover these results  imply the existence of magnetic domains in the samples, and therefore interactions between the individual magnetic moments of the porphyrins. In addition, the data is consistent with magnetic moments preferentially aligned, at very low T, in the molecules\rq{} plane, parallel to the graphene layer. This is evidenced by the strong  hysteretic odd component of the in-plane  magnetoresistance at low temperature in the graphene samples connected to normal electrodes (see supplementary materials). 
 By contrast, when graphene is connected to superconducting electrodes, the hysteresis occurs  also for perpendicular field. We explain this by the focusing and bending of the field lines by the superconducting electrodes, so that a perpendicular external field leads to a local field on the graphene sheet that contains both parallel and perpendicular components.  Finally, we have also checked that there is no magnetic hysteresis or asymmetry both with normal or superconducting electrodes in the absence of porphyrins (see supplementary materials).

The physics of magnetic impurities on graphene has led to a variety of interesting theoretical predictions specific to the band structure of graphene. Unlike metals, the exchange coupling in graphene is expected to  be controlled by gating \cite{Guinea12}. This effect has two important consequences. First, the amplitude of the magnetic moment of an individual impurity should strongly depend on gate voltage, causing a tunable Kondo effect. Depending on the relative energy of the impurity level with respect to the Dirac point, its magnetic state  could be highly asymmetric with gate voltage \cite{castro08} with  impurities which have a non zero magnetic moment for only one sign of gate voltage. Second, the long range magnetic  Rudderman, Kittel,  Kasuya,Yosida (RKKY) interactions  mediated by the carriers should also be gate  voltage dependent. Such interactions have been investigated theoretically and numerically  by several groups\cite{Guinea12,RKKY}.  Characteristic coupling energies in the Kelvin range are predicted between  spins 1/2 a few nanometers apart. The magnetic hysteresis we observe could then be explained by a spin glass type of order taking place at low temperature.  

Interestingly, it was recently shown  \cite{yao2014} that exchange interactions  in proximity induced superconducting   graphene are enhanced   on the scale of the coherence length compared to the normal state and acquire an antiferromagnetic sign in the vicinity of the Dirac point. Our results show that porphyrins coated graphene offers the possibility to explore this physics as well. Of course  extrapolating  these  theoretical findings on the exchange  coupling between localized spins on graphene to the case of Pt-porphyrins, where each spin is localized over the entire nanometer sized molecule is still a challenge \cite{Nickel}.

At this stage, let us summarize our understanding of the non-systematic occurrence of the gate-voltage polarity in the magnetism we report.
The gate-voltage polarity of magnetism is explained by the initial doping of graphene before porphyrins deposition, and the subsequent pinning of the HOMO or LUMO level of the Porphyrins upon coating and charge transfer. This leads to impurity states on average below the Dirac point for initially electron doped graphene (Nb sample), and above the Dirac point for initially hole-doped graphene (sample Ald1 and Ald2). As shown in \cite{castro08}, there is a limited gate range over which these impurity states are singly occupied and thus magnetic, and the range differs for both cases. The range nonetheless extends around the impurity levels energies, explaining why the magnetism should be more pronounced in positive gate voltage in the Nb sample, and negative gate voltage for the Ald1 and Ald2 samples. In contrast, samples with initially relatively broad   Dirac peaks centered close to zero gate voltage due to both electron and hole pockets (such as Au1 and Au2 samples), are bound to have impurity levels both above and below the Dirac point, in a relatively wide range, and consequently magnetism is expected for both positive and negative gate voltage.


In conclusion, this gate voltage-dependent long-range order of magnetic porphyrins mediated by graphene\rq{}s carriers is of great  potential interest  and motivates further investigations of different porphyrin species, including  metal free porphyrins or ones with a magnetic atom such as Fe or Co. Our findings  show that it is possible to control the number of magnetic porphyrins by their degree of ionisation. This suggests interesting possibilities  such as controling the magnetism by a gate voltage quench of graphene, or with a controlled concentration of charged defects.
 Finally we provide a new route for inducing gate-dependent long-range magnetism in graphene without destroying the sample mobility.

\section{Acknowledgements} We acknowledge ANR Supergraph and CNRS for funding, Juan Manuel Aguilar,  Alexei Chepelianskii, Richard Deblock,  Andrew Mayne, Claude Pasquier, Laurent Simon  and Alberto Zobelli for helpful discussions, and M. Feigelmann for suggesting that the suppressed proximity effect could be due to the molecules\rq{} magnetism.

\begin{figure}
      \centering
      \includegraphics[width=\textwidth]{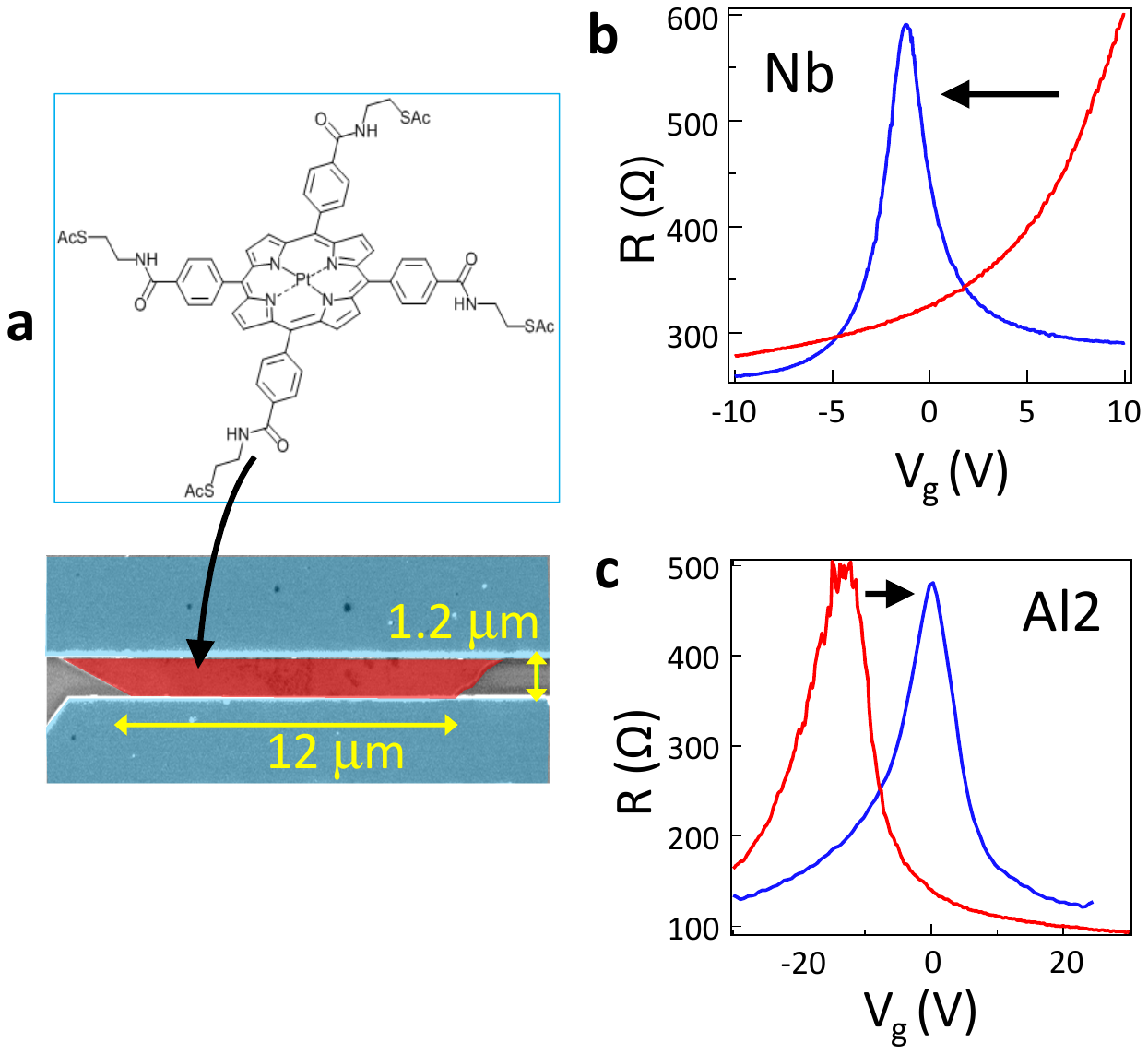}
      \caption{ \textbf{Charge transfer at room temperature} a: Electron microscope image of the graphene sample connected to  Pd/Nb electrodes, and  representation of the Pt-porphyrin. b and c: Gate dependence of resistance before and after  porphyrins  deposition for two samples, Nb (Pd/Nb electrodes) and Al2 (Ti/Al electrodes). Without porphyrins (red curves), the samples can be hole-doped (Dirac Point initially at positive gate voltage, (b)) or electron-doped (c). After grafting porphyrins, in both case, the Dirac point of graphene is brought to  practically zero: graphene has become neutral. This implies that charge transfer occurs between graphene and porphyrins and the molecules can be donors (b) as well as acceptors (c) of electrons. We believe that this neutralization  of the charged defects in graphene occurs because it reduces  the electrostatic energy of the system   thanks to a better spatial  confinement than when charged impurities are screened by graphene\rq{}s carriers (since that screening occurs over larger distances, of the order of the Thomas-Fermi screening length (10 to 30nm)). This charge transfer is accompanied by gate voltage hysteresis and relaxation effects (see Fig.2). 
			}
      \label{Fig1}
\end{figure}

\begin{figure}
      \centering
      \includegraphics[width=\textwidth]{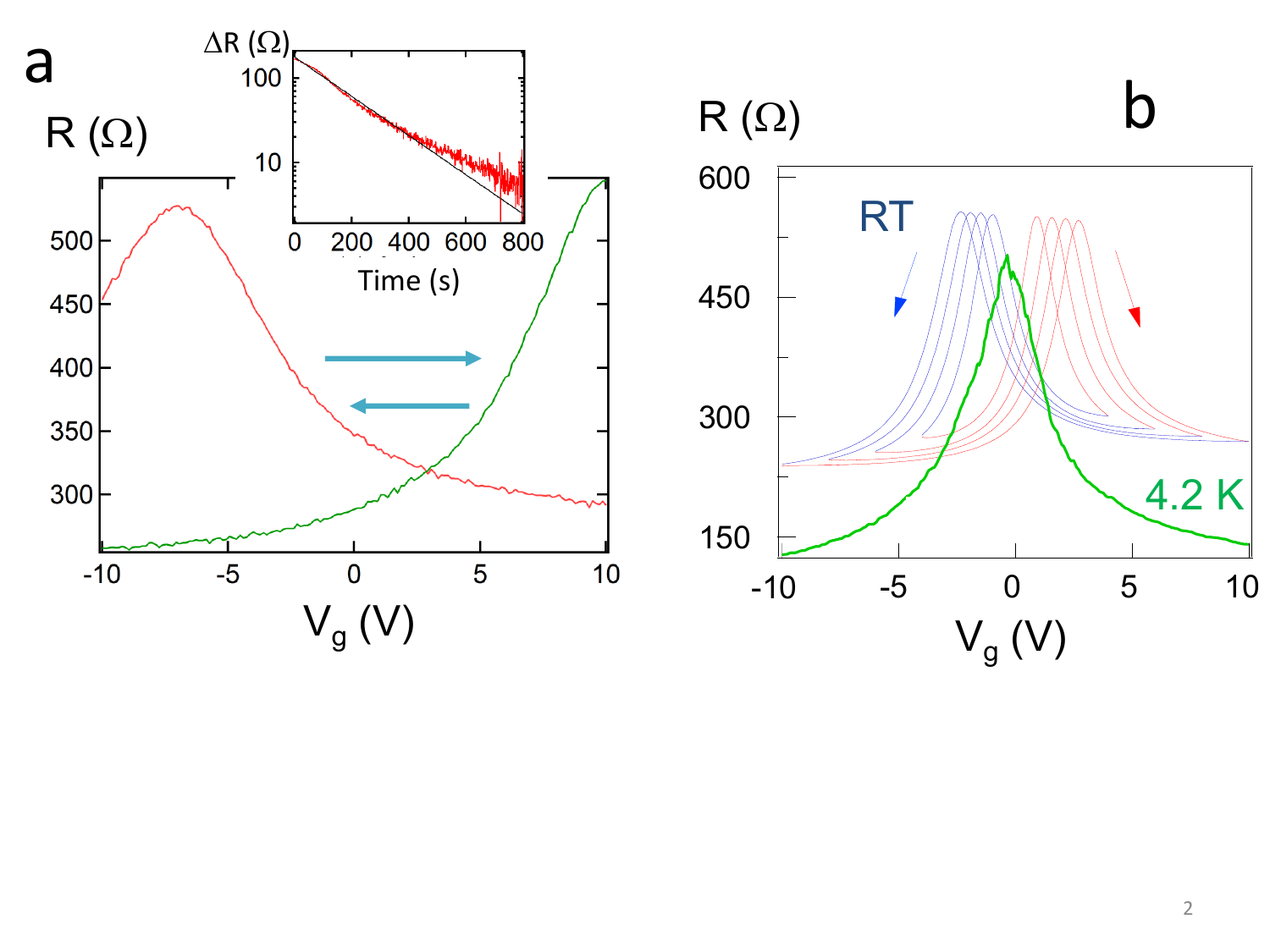}
      \caption{ \textbf{Hysteresis in gate dependence at room temperature of the porphyrin-coated Pd/Nb sample}  (a) Modification of the Dirac point position after keeping the sample for one hour at $ V_G = \pm 10 V $. Inset: Time evolution of the sample resistance at fixed gate voltage after a rapid gate shift from -10 to +10 V. The slow relaxation can be fitted by $ \Delta R = R_0 e^{-t/t_0} $ with $R_0 = 165\Omega$ and $t_0$ = 187 s. (b)  Gate dependent resistance at
room temperature for different excursions. One can note the hysteresis, which amplitude
depends on the gate bias excursion in the presence of porphyrins. Bottom (green) curve: Gate-dependent
resistance measured at 4.2 K. The hysteresis is absent at low temperature. The vertical shift to lower resistance is due to the superconductivity of the Pd/Nb electrodes. 
			}
      \label{Fig2}
\end{figure}

\begin{figure}
      \centering
      \includegraphics[width=0.7\textwidth]{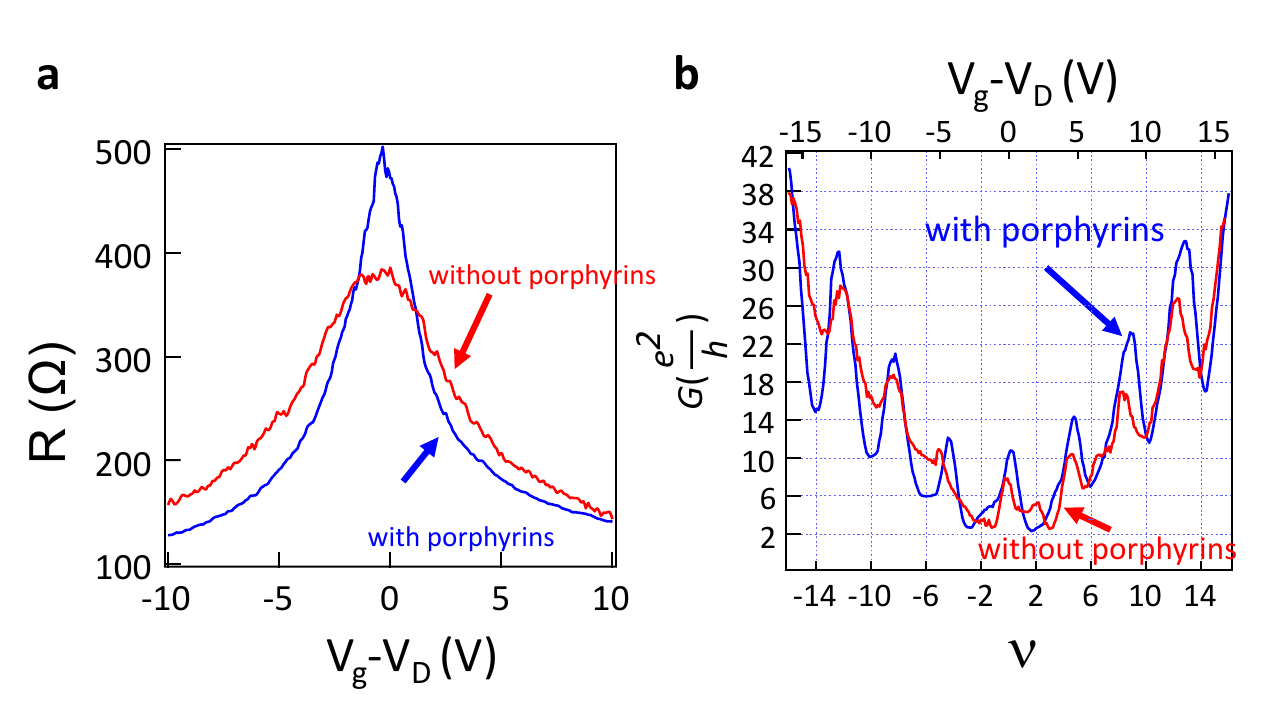}
      \caption{\textbf{Improvement of the sample quality after porphyrins deposition (Pd/Nb sample)}. a: Zero magnetic field gate voltage sweep of two wire resistance at 4.2 K, before (red) and after (blue) deposition of Pt-porphyrins. The origin is taken at the Dirac point $ V_D $ for both curves. The sharper curve with a higher Dirac peak demonstrates that the sample quality has improved thanks to porphyrin coating, with a mobility increase of 20\%. b: Two-wire measurement of the resistance in the quantum Hall regime, at 100 mK and in a perpendicular magnetic field of B = 5 T (for which the Nb electrodes are non superconducting), before (red curve) and after (blue curve) grafting. The gate voltage is expressed in terms of the filling factor $ \nu = n_c \phi_0 /B $, with the charge density $n_c$ computed assuming that only the backgate charges the graphene. The Hall plateaus (visible even in a two wire measurement) are better defined, confirming the higher sample quality. The gate voltage position of the plateaus is unchanged, demonstrating that the charge transfer between porphyrins and graphene is gate independent, and fixed at low temperature.}
      \label{Fig3}
\end{figure}
\begin{figure}
      \centering
      \includegraphics[width=\textwidth]{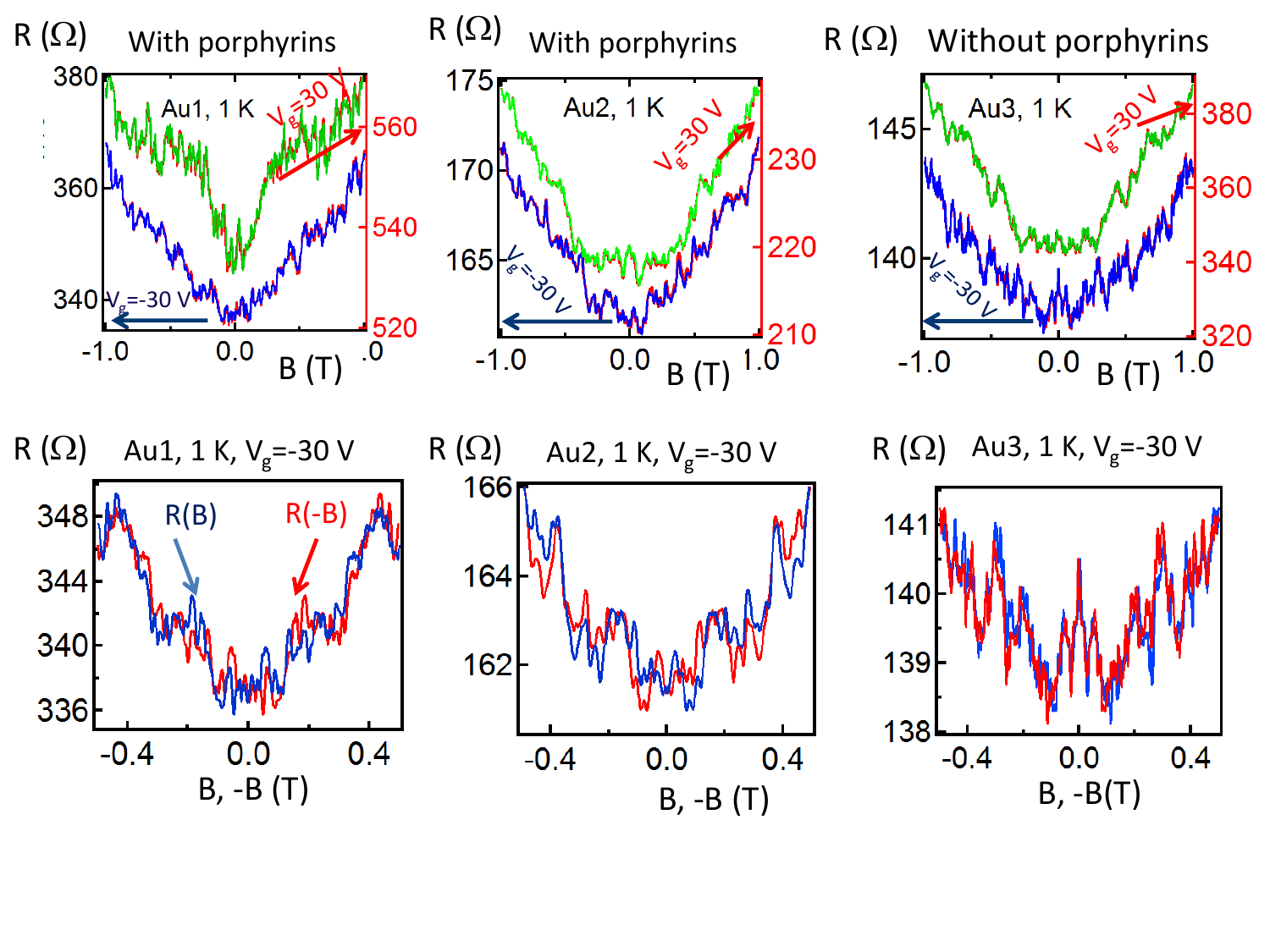}
      \caption{ \textbf{Field asymmetry of the perpendicular magnetoresistance  in porphyrins coated samples  with Ti/Au electrodes at 1 K}. Out-of-plane magnetoresistance of three samples on the same chip is presented. Au1 and Au2 were coated with Pt-porphyrins, and Au3 was not. The resistance fluctuates with magnetic field in a reproducible way for all three samples (overlapping traces correspond to successive field sweeps).  We find that the magnetoresistance curves of the coated samples are asymmetric.
             This asymmetry   is  better illustrated by plotting both the R(B) and R(-B) curves, see bottom panels. The asymmetry is clear for the samples coated with Porphyrins,  Au1 and Au2, but not for the uncoated sample Au3. } 
      \label{Fig4}
\end{figure}

\begin{figure}[h]
      \centering
      \includegraphics[width=0.8\textwidth]{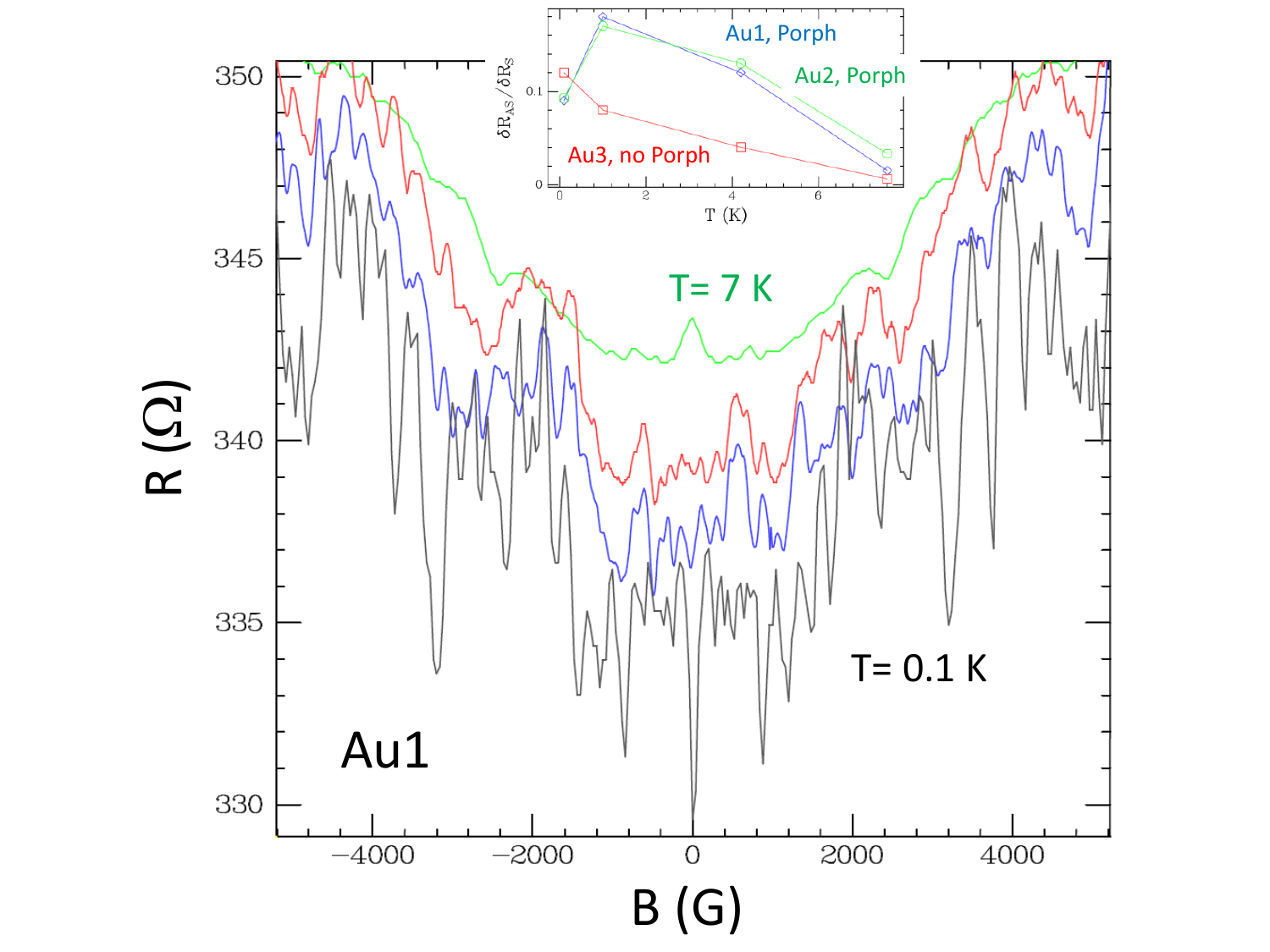}
           \caption{\textbf{T dependence of asymmetry of resistance fluctuations in a perpendicular field for samples with Ti/Au electrodes.} Resistance fluctuations in a perpendicular field for coated sample Au1, at 100 mK, 1 K, 4.2 K and 7 K. The inset shows the asymmetry of the fluctuations (variance of the (R(B)-R(-B)) curve, renormalized by the symmetric fluctuations (variance of R(B)+R(-B)) for all three samples (averaged over 4 curves for each sample), at $V_g=-30 V$. The asymmetry of the coated samples Au1 and Au2 is more than twice that of the uncoated sample Au3 at 1 K and 4.2 K.  The relative asymmetry of the coated samples is maximal at 1 K, when the moments are perpendicular to the graphene plane, and phase coherence is still achieved throughout the sample, so that the magnetoresistance fluctuations are sensitive to the magnetic moments. It is still visible at 4.2 K.  The loss of asymmetry at 7 K where mesoscopic fluctuations  are reduced but  still can be detected, can be attributed to the  loss of long range magnetic order.  The loss of asymmetry at 100mK for the coated samples is explained by a moment in plane at low T. The slight low temperature increase of the asymmetry observed for the uncoated Au3 sample is due to the increase of mesoscopic telegraphic noise at low temperature observed on all graphene samples. }
     \label{Fig5}
\end{figure}

\begin{figure}
      \centering
      \includegraphics[width=0.7\textwidth]{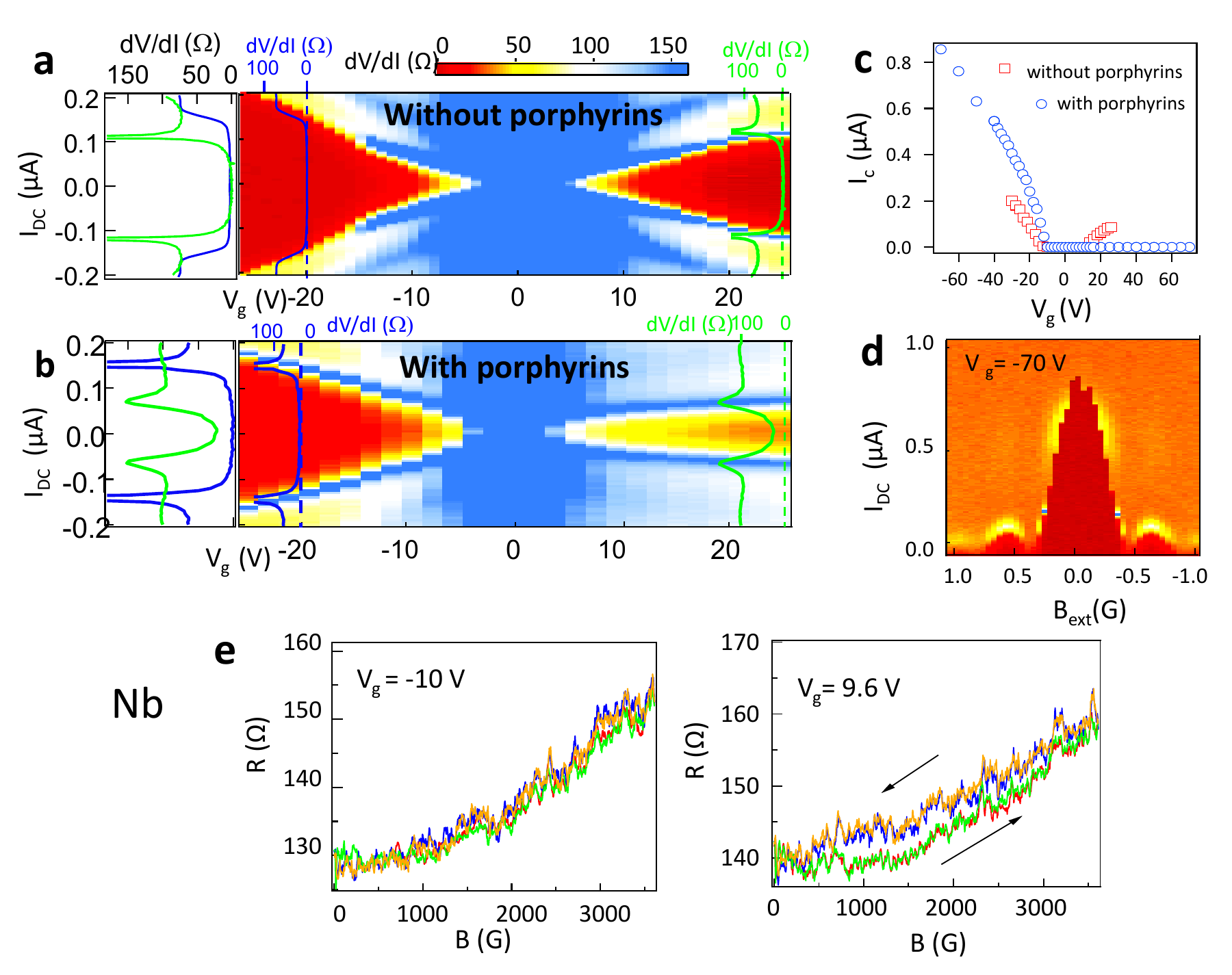}
      \caption{\textbf{Change from bipolar to unipolar induced superconductivity upon deposition of Pt-porphyrin on a sample with Pd/Nb electrodes.}(a,b) Color-coded differential resistance as a function of dc current (y axis) and gate voltage (x axis), measured with a small ac current added to the dc current. The dark red regions correspond to regions of zero differential resistance where a Josephson supercurrent runs through the S/graphene/S junction. Whereas the Josephson effect occurs symmetrically about the Dirac point on the pristine, uncoated sample (\textbf{a}, T=200 mK), it only occurs on the hole doped side (negative $V_G$ ) on the sample covered with porphyrins (\textbf{b}, T = 100 mK). The curves on the left of the color plots are the differential resistance curves as a function of dc current,  at gate voltages symmetric with respect the Dirac point.  \textbf{c}: Change from bipolarity to unipolarity upon coating, revealed by the variations with gate voltage of the critical current $I_c$, i.e. highest dc current for which the differential resistance is zero. Before (red squares, bipolar) and after (blue circles, unipolar) porphyrin deposition.  \textbf{d}: Differential resistance as a function of dc current and external magnetic field (perpendicular to the graphene sheet), for the graphene with porphyrin molecules, at $V_G$ = -70 V revealing a Fraunhofer pattern. (e) Hysteresis in the graphene\rq{}s magnetoresistance after porphyrin deposition, at 100mK, with a field direction perpendicular to the graphene plane. The hysteresis is quite large at positive gate voltage (electron doping, right curves) and negligible (at least an order of magnitude smaller) for hole doping (left curves), confirming the existence of a magnetic order that suppresses the supercurrent for electron doping. }
      \label{Fig6}
\end{figure}

\begin{figure}
      \centering
      \includegraphics[width=0.6\textwidth]{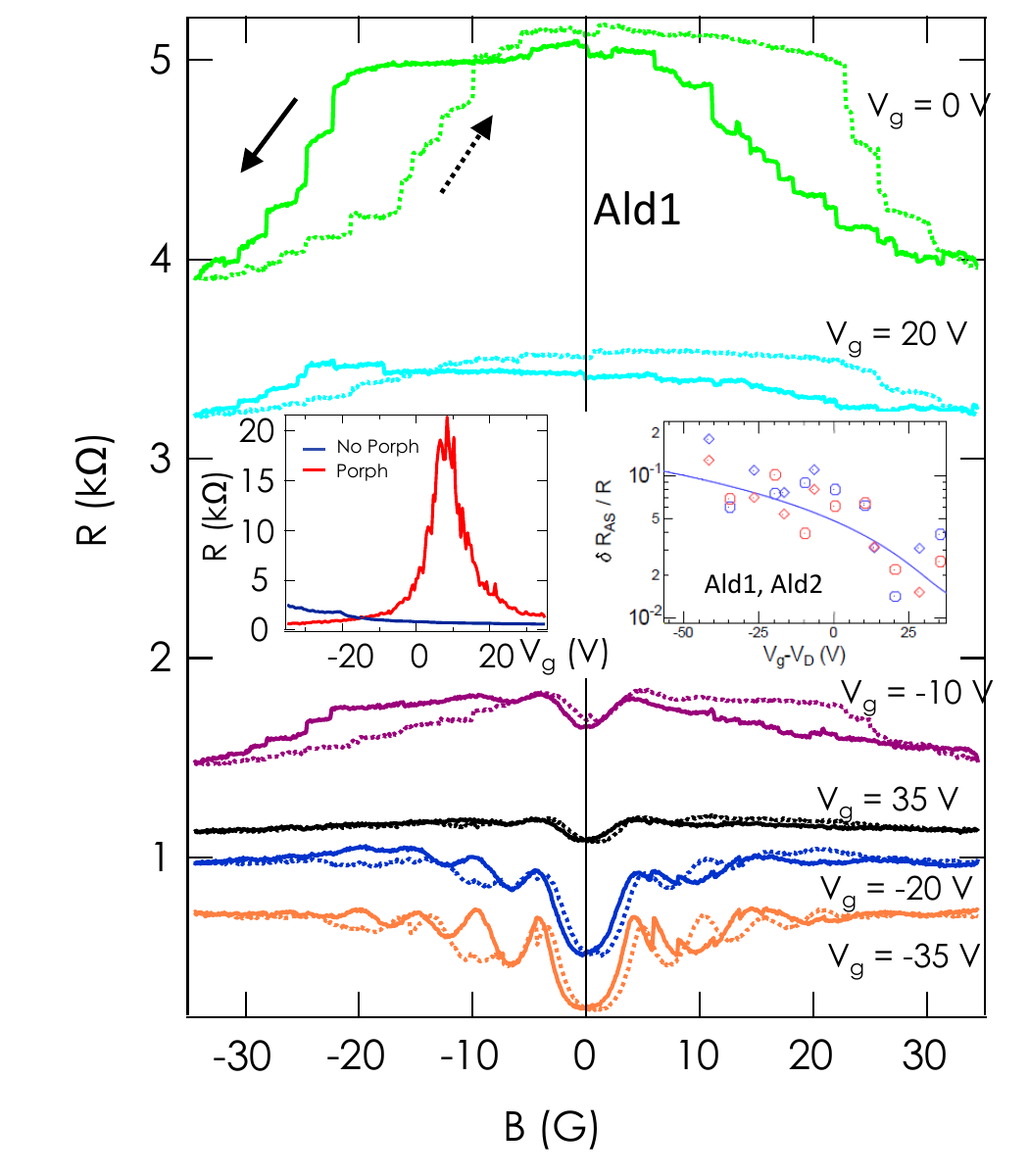}
      \caption{\textbf{ Proximity induced superconductivity  (Ti/Al contacts) in  a  low mobility long junction sample  coated with porphyrins}. Magnetoresistance   of sample  Ald1 at different gate voltages. Up and down magnetic field sweeps are respectively dashed and solid lines, the sweep directions are indicated by arrows. The hysteresis at low doping is accompanied by many jumps corresponding to strongly correlated magnetic domains.  Left inset: gate dependence before (blue) and after (red) porphyrin deposition, showing the striking neutralization power of porphyrin molecules on graphene and,  in this case,  the property of porphyrins to act as electron acceptors. Right inset : Gate voltage dependence of the antisymmetric component of the magnetoresistance $\delta R_{AS}$ rescaled to the mean resistance R, of the samples Ald1 (diamonds) and Ald2 (circles). Red and blue points correspond to  up and down magnetic field sweeps respectively.} 
      \label{Fig:chapPorph_G92disorder}
\end{figure}

\newpage
\begin{table}[bp] 

\begin{tabular}{|c|c|c|c|c|c|}
\hline
Sample &  Contact & Length ($\mu m$)& width ($\mu m$) &$\mu ~ (cm^2V^{-1}s^{-1})$  & $\Delta V_D ~(V)$ \\ \hline
$Au1$  & Ti/Au& 0.8 & 3.5 & 2000& -10  \\ \hline
$Au2$ & Ti/Au& 0.7 & 8  & 1400 & $\leq - 20$ \\ \hline

$Au3$  & Ti/Au& 0.8 & 3.5 & 1600 & no porph.   \\ \hline
$Nb$& Pd/Nb&1.2&12& 5000 & -10\\ \hline
$Al1$&Ti/Al&0.5 & 3.4& 3000&no porph.\\ \hline
$Al2$&Ti/Al&0.45& 4 &2000&  15 \\ \hline
$Al3$&Ti/Al&0.5& 4 &2000&  15 \\ \hline
$Ald1$&Ti/Al&0.4& 2.6 &150&  $\geq 40 $\\ \hline

$Ald2$&Ti/Al&0.6& 4 & $\leq 100$ &  $\geq 30$ \\ \hline
\end{tabular}
\caption{ Principal characteristics of the samples cooled down after deposition of porphyrins. (Except control samples Au3 and Al1 for which no porphyrins were deposited). The mobility $\mu$  was  measured at room temperature  near the Dirac point before porphyrin deposition. $\Delta V_D$ is the shift in gate voltage  of the Dirac point after deposition of the porphyrins. This number gives an indication of the concentration of ionized porphyrins: A shift of 10 to 40 V in Dirac point upon porphyrin deposition corresponds to a transfer of $1~to~ 4~10^{12}$ charges per $cm^2$, implying that 4 to 15\% of the porphyrin molecules in contact with graphene ionize.  }

\label{table}
\end{table}

\begin{thebibliography}{99}
\bibitem{Fraunhoffer}A  flux quantum through the area of the sample ($1.2 \times 12 \mu m^2 $ corresponds to a field of 3 Gauss,  which is much larger than the value of 0.4 Gauss measured experimentally. We understand this effect (also reported in many other papers on the proximity effect in 2D materials) as due to the focusing of field lines between the 2 massive Nb superconducting electrodes whose width is equal to 5 microns.
\end{thebibliography}
\end{document}